\documentclass[%
 reprint,
superscriptaddress,
 amsmath,amssymb,
 aps,
10pt]{revtex4-2}

\usepackage{graphicx}
\usepackage{dcolumn}
\usepackage{bm}

\usepackage{subcaption}

\usepackage{scalerel}

\usepackage[dvipsnames]{xcolor}

\begin{document}

\pagenumbering{gobble}


\title{A High-Dimensional Lab-Frame Model of Markovian Dynamics in Alanine Dipeptide}





\author{J. M. Hall}
\email{jhall4@uoregon.edu, jmonroe.hall4@gmail.com}

 \affiliation{%
Department of Physics, University of Oregon, Eugene, Oregon 97403, USA}%

\author{M. G. Guenza}%
 \email{mguenza@uoregon.edu}

\affiliation{%
 Department of Chemistry and Biochemistry, University of Oregon, Eugene, Oregon 97403, USA
}%

\date{\today}

\begin{abstract}
We present a high-dimensional coarse-grained model of alanine dipeptide that explicitly incorporates global translation and rotation alongside internal conformational dynamics. Using atomistic molecular dynamics trajectories, we construct a 10-bead Markovian Langevin model with a 24-dimensional internal potential of mean force and a full matrix of site-specific friction coefficients. The equilibrium landscape is represented by a Gaussian mixture model, while a rotational-entropy correction accounts for the coupling between internal conformation and global molecular motion. Friction coefficients are obtained using a generalized Einstein relation based on time-correlation functions. We show that the rotational-entropy correction is essential for obtaining a physically admissible friction matrix and significantly affects the inferred dynamics. Without fitting to coarse-grained dynamical observables, the resulting model reproduces a broad range of atomistic auto- and cross-correlation functions. The result is a compact and interpretable dynamical
representation of the entire peptide at the chosen CG resolution, which explicitly preserves distinctions
between energetic and dissipative coupling over a wide range of motions, including those between
conformational fluctuations and global translational and rotational diffusion.
\end{abstract}

\maketitle


Coarse-grained (CG) models describe complex physical systems in terms of a small number of coordinates, aiming to statistically reproduce the dynamics of these coordinates while treating the remaining degrees of freedom only implicitly. For example, a protein in solvent may be described by the motion of the alpha carbons in each amino acid, and the solvent composed of water molecules modeled only implicitly by a drag force and a fluctuating noise, as in a Langevin equation \cite{beyerleGuenza2021}. Such models generally have the advantages of analytical simplicity and computational accessibility relative to their fine-grained counterparts. Achieving their construction without a loss in accuracy is an area of ongoing research.

In particular, there has been much recent activity in the development of parametrization methods for Markovian and Generalized Langevin Equations consistent with the underlying Mori-Zwanzig projection operator formalism. These models have the potential to be highly accurate, and their formal grounding in statistical-mechanical theory allows for well-defined, controlled approximations.

A common strategy is therefore to organize the projected dynamics around one or a small number of internal reaction coordinates. Such representations are powerful when
a particular process is known in advance and can be characterized by an appropriate collective variable, but they are also limited in the questions that the resulting model can address. A one-dimensional GLE can describe the kinetics of a selected transition with considerable fidelity, yet the dynamics of a complex molecule are not generally organized around a single process. Conformational modes can be coupled to one another, and a clear separation between relevant and irrelevant degrees of freedom is not guaranteed. Furthermore, the network of couplings itself may be an object of interest, determining how fluctuations propagate through a molecule. Coupling between global translation, rotation and internal fluctuations may be relevant during certain binding events, and is intrinsically absent from low-dimensional models. Even when good reaction coordinates exist, when they are not known \textit{a priori}, a higher-dimensional representation may serve as a fruitful intermediate tool for exploratory work.

A qualitatively complete dynamical model must include not only conformational fluctuations, but translation and rotation in the lab frame. Two developments are required to construct such a representation.

First, the friction associated with a molecular coarse-graining must be determined in the presence of both bound internal coordinates and unbounded translational motion. We recently developed a generalized Einstein relation that obtains Markovian friction coefficients directly from observable time-correlation functions. An important feature of this relation is the freedom to choose the correlations used in the estimator, which permits its application to systems containing coordinates with different dynamical symmetries by choosing correlations that possess a numerically stable eigenvalue structure.

Second, the equilibrium conformational distribution must be treated consistently when internal coordinates are defined in a body-fixed frame but the resulting model is intended to describe a freely translating and rotating molecule in the laboratory frame. We previously showed that direct Boltzmann inversion of the body-fixed configurational distribution does not give the appropriate laboratory-frame potential because eliminating rotational degrees of freedom from the distribution introduces a conformation-dependent rotational entropy. Correcting for this contribution recovers the potential of mean force governing the corresponding laboratory-frame dynamics.

In this work we apply both of these methods for the first time to the trajectory of a real molecule, alanine dipeptide, obtained from an unbiased Molecular Dynamics (MD) simulation with explicit solvent. We construct a site-specific 10-bead model, with an internal 24-dimensional energy landscape and a $10\times10$ matrix of Markovian friction coefficients. We demonstrate the stability of the Markovian plateau obtained from the Generalized Einstein Relation as described in \cite{frictionHallGuenza}, and we demonstrate the necessity of the rotational entropy corrections described in \cite{rotEntropyHallGuenza}. Because no enhanced sampling is used, certain barriers are not ideally sampled and there is much room for improvement in the model at long timescales in coordinates relevant to a rare metastable state. Even so, we find the qualitative agreement in the time correlation functions between the MD and CG models remarkable.

Our main goal in this letter is to demonstrate the feasibility of extending data-driven Mori-Zwanzig-type models to high-dimensional spaces, particularly those which explicitly include translational and rotational symmetry. We will demonstrate the principles involved and how our methods permit their construction, and then discuss the potential merits of such models and the remaining practical obstacles to wider application.

\section{Molecular Dynamics Simulation}

Alanine dipeptide in TIP3P water solvent \cite{tip3p} was simulated in GROMACS 2020.4 \cite{gromacs2020} using the AMBER99SB-ILDN force field \cite{amberff} with Particle-Mesh Ewald electrostatics and a 1nm non-bonded cutoff. A dodecahedral periodic box was used with unit cell length of 32.326\r{A}. The box was filled with 771 water molecules, along with one pair of Na$^{+}$ and Cl$^{-}$ ions to achieve a salt concentration of approximately 0.072M. The peptide was relaxed in the presence of the solvent and ions using the GROMACS steepest-descent algorithm. Two 1ns equilibration runs were performed in the NPT and NVT ensembles respectively at a timestep of 2fs. Both were performed at a temperature of 298K using the v-rescale thermostat and the latter at a pressure of 1atm using a Parinello-Rahman barostat. For each of ten replicas, we generated 490ns production runs using the v-rescale thermostat, again at 298K, with a timestep of 1fs. These high-resolution runs were used to calculate the CG friction coefficients as described below. To resolve the equilibrium landscape, 98 production runs were generated for 490ns each under the same conditions at a lower 10fs resolution, for a total simulation time of 48.02$\mu$s. All production runs were preceded by 10ns of discarded relaxation under the same conditions. Hydrogen bond lengths were constrained in all simulations using the SHAKE algorithm. Simulations were performed on the Talapas High-Performance Computing cluster at the University of Oregon.

\section{Distribution of Internal Coordinates}

Alanine dipeptide consists of 22 atoms, 12 of which are hydrogen atoms. In this treatment, every hydrogen atom was grouped with its heavy bonded partner, and the center-of-mass of each resulting group defined a CG site. This yields a $N=10$ bead model of alanine dipeptide, with 30 total Cartesian degrees of freedom.

In order to retain translational and rotational symmetry, the potential of mean force for an $N$-bead CG model must depend only on the $3N-6$ (in this case 24) internal degrees of freedom of the system.

Let $\vec{r}$ be the lab-frame positions of the $N$ beads. Then let $\vec{r}_{B}$ be the body-fixed positions as defined by the Eckart frame with reference structure $\vec{r}_{0}$. The reference structure was defined by the average body-fixed structure (generated using the initial conformation as a dummy reference) such that fluctuations of the internal displacements are zero-centered.

As discussed in \cite{rotEntropyHallGuenza}, the distribution of internal coordinates is affected by the rotational entropy associated with the global diffusion. In that work, it was shown that the internal shape coordinates, and the body-fixed frame used to calculate them, may be chosen carefully to greatly simplify the correction for these effects. This permits extraction of the potential of mean force from the free energy landscape. As such, we will use the same scheme for the internal coordinates deployed in \cite{rotEntropyHallGuenza} and do not repeat the details here.

To construct the CG force, we will need to calculate the gradient of the probability distribution $\mathcal{P}(\vec{q})$ for any given $\vec{q}$. This requires an analytical fit to the observed $\mathcal{P}(\vec{q})$. Due to the high-dimensional nature of the distribution, any methods involving the explicit construction of a histogram are impractical.

We instead represent $P(\vec{q})$ analytically as a Gaussian mixture model (GMM),
\begin{equation}
P(\vec{q})
=
\sum_{k=1}^{n} w_k\,
\mathcal N(\vec{q};\bm{\mu}_k,\bm{\Sigma}_k),
\end{equation}
where $w_k$, $\bm{\mu}_k$, and $\bm{\Sigma}_k$ are the weight, mean, and covariance matrix of mixture component $k$.

We use the Expectation-Maximization (EM) algorithm to construct an approximate Maximum Likelihood Estimation for the parameters of the GMM directly from the trajectory data.

The number of gaussians $n$ is a hyperparameter which must be chosen before the EM algorithm is initiated, and so to choose a good value for $n$ we perform many fits with different values of $n$ and then check a validation measure to discern the best fit. Simply increasing $n$ almost always improves the final value of the likelihood, even when overfitting is occurring. As another validation measure, then, we calculate the error in satisfying the generalized equipartition theorem

\begin{equation}
\langle q_{i}\frac{\partial \mathcal{H}}{\partial q_{j}}\rangle = k_{B}T\delta_{ij}
\end{equation}

over the original trajectory. Since the body-fixed coordinate system is obtained from the lab frame coordinates via a canonical transformation, the above equation implies that

\begin{equation}
-k_{B}T\scaleleftright{\langle}{q_{i}\frac{1}{\mathcal{P}(\vec{q})}\frac{\partial \mathcal{P}(\vec{q})}{\partial q_{j}}}{\rangle} = k_{B}T\delta_{ij}
\end{equation}

which can be checked using the fit for $\mathcal{P}(\vec{q})$.

\begin{figure}[t]
\includegraphics[width = 8.6cm]{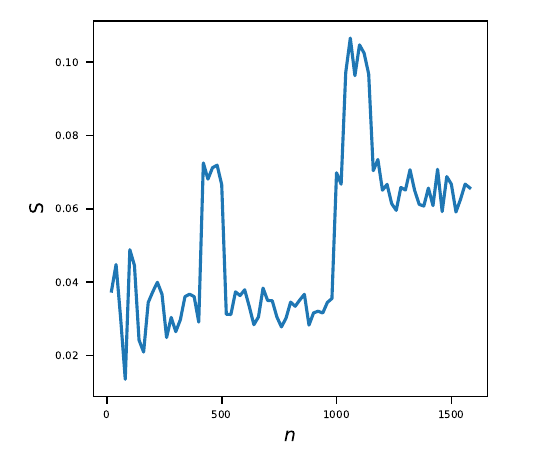}
\caption{\label{fig:scoring} The equipartition scores $S$ obtained from Gaussian Mixture Models of varying component number $n$. The landscape selected for the CG model was $n=960$.}
\end{figure}

\begin{figure}[t]
\includegraphics[width = 8.6cm]{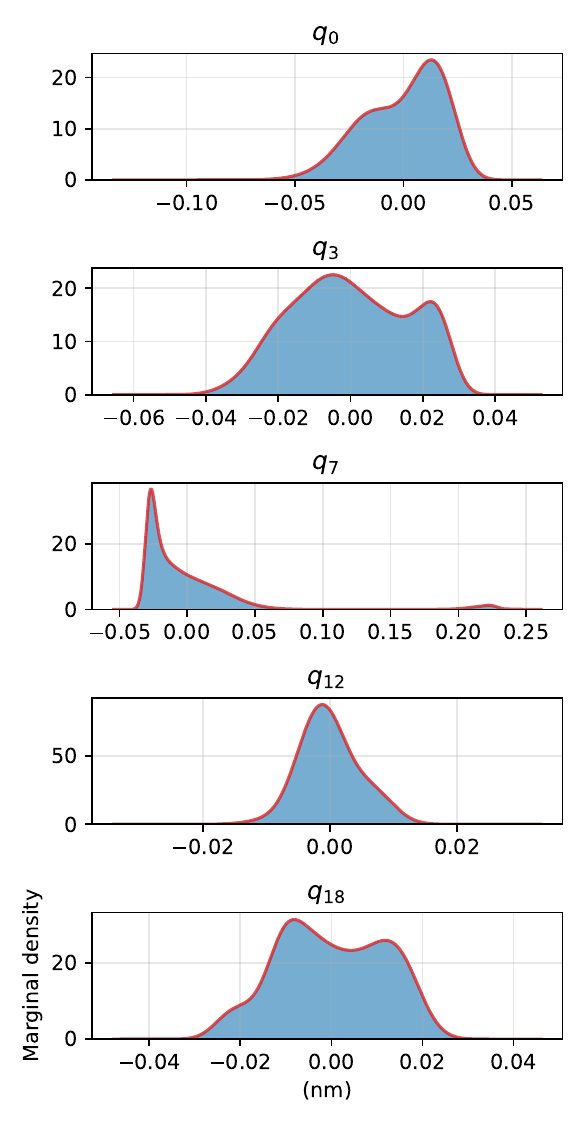}
\caption{\label{fig:aladi_prob_fit} Histograms (transparent) of shape coordinates obtained from the MD simulation of alanine dipeptide, overlaid with the corresponding marginal distributions predicted by a 960-component Gaussian Mixture Model (GMM) fit (solid lines).}
\end{figure}

We define a score $\mathcal{S}$ such that

\begin{equation}
\mathcal{S}:= \bigg| -k_{B}T\scaleleftright{\langle}{\vec{q}\frac{1}{\mathcal{P}(\vec{q})}\frac{\partial \mathcal{P}(\vec{q})}{\partial \vec{q}}^{T}}{\rangle} - k_{B}T\mathbb{I} \bigg|
\end{equation}

where the vertical bars represent the Frobenius norm. A particular fit is considered good when this score is small as calculated on the observed trajectory. We note that it is trivially equal to zero, on average, if the samples are instead drawn from the GMM itself. We have chosen this measure heuristically from experience, as force fields with a high $\mathcal{S}$ score almost invariably yield friction matrices with unphysical negative eigenvalues.

In Fig. 1 we plot the scores of a range of fits, with the number $n$ of gaussian components ranging from 20 to 1600. Below n~1000, most models yield similar scores. At higher values of $n$ the score degrades even as the likelihood increases, possibly due to overfitting. To balance these conditions, our strategy was to choose the local minimum with the greatest value of $n$ before this jump in $S$.

In Fig. 2 we plot histograms of select shape coordinates from the MD simulation, together with the curves of marginal distributions calculated from the $n=960$ GMM. The visual agreement in regions of high probability is very good across all $24$ internal coordinates. A few hundred gaussians seem largely adequate to describe the basins of even a high-dimensional molecular landscape.

\section{The Potential of Mean Force and Rotational Entropy}

The probability distribution of body-fixed shape coordinates cannot be converted directly into the laboratory-frame intramolecular potential through a naive Boltzmann inversion. As shown in our previous work\cite{rotEntropyHallGuenza}, elimination of the global rotational degrees of freedom introduces a conformation-dependent rotational entropy into $P(\vec{q})$. For the rectilinear Eckart-frame coordinates used here, the corresponding potential of mean force is therefore given by
\begin{equation}
U(\vec{q})=
-k_B T\ln P(\vec{q})
+
k_B T
\ln\left[
A\sqrt{|I^*(\vec{q})|}
\right],
\label{eq:pmf}
\end{equation}
where $I^*(\vec{q})$ is the generalized moment-of-inertia tensor and $A$ is a coordinate-independent normalization factor.

The generalized force associated with shape coordinate $q_i$ is consequently
\begin{equation}
F_i(\vec{q})
=
-\frac{\partial U}{\partial q_i}
=
k_B T\frac{\partial\ln P(\vec{q})}{\partial q_i}
-
\frac{k_B T}{2}
\mathrm{Tr}
\left[
I^{*-1}
\frac{\partial I^*}{\partial q_i}
\right].
\label{eq:force}
\end{equation}
The first term is obtained analytically from the GMM, whereas the second accounts for and removes the shape-dependent rotational contribution to the body-fixed free-energy landscape. The definition and derivation of $I^*(\vec{q})$, together with the procedure used to transform the 24 generalized forces into the full set of 30 laboratory-frame Cartesian forces, are provided in ~\cite{rotEntropyHallGuenza}.

For comparison, subsequent friction calculations were performed using both the complete rotationally corrected force and the uncorrected force obtained by omitting the second term in Eq.~\eqref{eq:force}. This allows the influence of the rotational-entropy correction on the inferred dynamics of a realistic solvated molecule to be evaluated directly.

\section{Friction Coefficients}

In a many-bead CG model of a molecule, the site-specific friction coefficients are responsible for the kinetics of both the internal shape changes and the overall translational and rotational diffusion of the molecule. In previous work \cite{frictionHallGuenza} we have shown how Markovian friction coefficients simultaneously consistent with all of these behaviors on long timescales can be obtained in a manner consistent with an underlying memory function. This method made use of a Generalized Einstein Relation (GER)

\begin{equation}
\zeta_{g}(t) = \left[ MC^{gv}(0) - MC^{gv}(t) + \mathcal{F}^{gU'}(t) \right] {\mathcal{D}^{gv^{-1}}}(t),\label{eq:zeta_g_formula}
\end{equation}

where

\begin{equation}
C^{xy}_{ij}(t) := \langle\vec{x}_{i}(0)\cdot\vec{y}_{j}(t)\rangle
\label{eq:correlation_def}
\end{equation}

\begin{equation}
\mathcal{F}^{vU'}(t) =-\int_{0}^{t}d\tau\:C^{vU'}(\tau) \label{eq:F_def}
\end{equation}

\begin{equation}
\mathcal{D}^{vv}(t) := \int_{0}^{t}d\tau \:C^{vv}(\tau) \ , \label{eq:D_def}
\end{equation}

In this relation, the long-time limit of $\zeta_{g}$ yields the Markovian friction coefficients for any choice of $\vec{g}$, so that

\begin{equation}
\zeta_{g}(\infty) = G(\infty) = \zeta\label{eq:zeta_g_limit}.
\end{equation}

We adopt the hybrid choice of g previously shown to provide numerical stability in systems containing both bound internal coordinates and unbounded center-of-mass diffusion. The generalized force $-U'(\vec{q})$ entering the friction calculation is evaluated from the 24-dimensional GMM using Eq. \eqref{eq:force}.

\begin{figure}[t]
\includegraphics[width = 8.6cm]{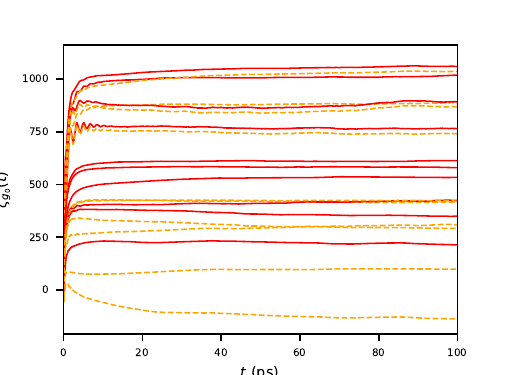}
\caption{\label{fig:friction_diag} The diagonal components of the friction obtained using the entropically corrected (solid) and uncorrected (dashed) force fields.}
\end{figure}

\begin{figure}[t]
\includegraphics[width = 8.6cm]{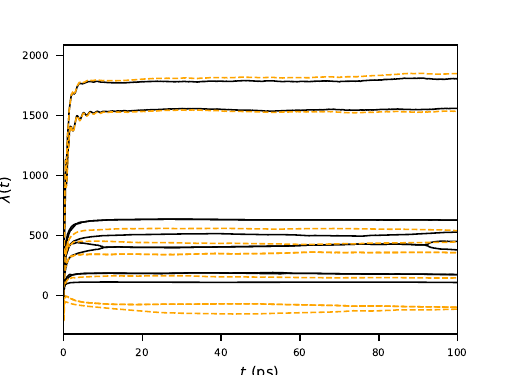}
\caption{\label{fig:friction_eigs} The eigenvalues of the friction obtained using the entropically corrected (solid) and uncorrected (dashed) force fields.}
\end{figure}

\begin{figure}[t]

\begin{subfigure}{8.6cm}
\includegraphics[width = 8.6cm]{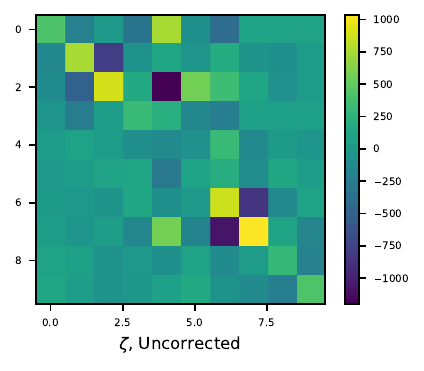} \end{subfigure}

\vspace{5mm}

\begin{subfigure}{8.6cm}
\includegraphics[width = 8.6cm]{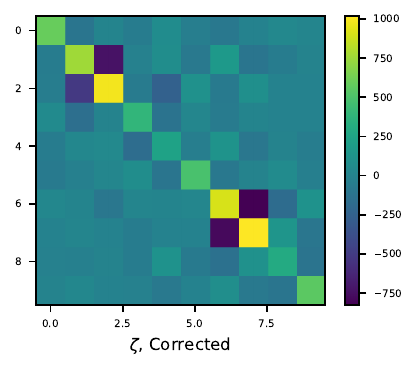}
\end{subfigure}
\caption{\label{fig:friction_mat} The friction matrices drawn from $\zeta_{g}(t)$ at $t=$100ps using the entropically uncorrected (top) and corrected (bottom) force fields.}
\end{figure}

In Fig. \ref{fig:friction_diag}, we show the resulting $\zeta_{g}(t)$ using the rotational entropy-corrected and uncorrected force fields. We see that the approach does indeed yield stable plateau values in both cases. However, as shown in Fig. \ref{fig:friction_eigs}, the uncorrected forces yield a Markov friction matrix with negative eigenvalues. This is not usable in a CG model as the trajectory would accumulate energy rather than dissipate it. Of note, it is the low-friction modes that are most significantly perturbed by the rotational effects, while the greatest eigenvalues are largely unaltered. The corrected force field, on the other hand, yields a positive definite friction matrix. Additionally, the off-diagonal components of the friction matrix are far more symmetric when using the corrected forces. In Fig. \ref{fig:friction_mat} we can see that the corrected force field yields a far more structured friction matrix.

These results are striking, and indicate significant effects from rotational-vibrational coupling in real molecules. The correction itself is a very small fraction of the instantaneous force, yet its effects alter the force TCF at long times and affect many eigenvalues of the friction matrix significantly. Our previous work only demonstrated the necessity of this correction in the case of a particular toy model. Here we see that they remain crucial in the case of a small peptide in explicit solvent.

\section{Coarse-Grained Simulation}

With a force field and friction coefficients in hand, a Langevin equation for the 10-bead model of alanine dipeptide can be constructed and simulated in the lab-frame.

The Markovian Langevin equation of the CG model is given by

\begin{equation}
\begin{aligned}
\dot{\vec{r}}(t) &= M^{-1}\vec{p}(t)\\
\dot{\vec{p}}(t) &= -\frac{\partial V(\vec{q}(t))}{\partial \vec{r}} - \zeta\vec{p}(t) + f^{R}(t)
\end{aligned}
\end{equation}

The cross-bead correlations in the noise force are obtained from the second Fluctuation-Dissipation Theorem via the matrix square-root of $\zeta$ such that

\begin{equation}
\langle f^{R}_{i\alpha}(t)f^{R}_{j\beta}(t')\rangle = 2k_{B}T\zeta_{ij}\delta_{\alpha\beta}\delta(t-t').
\end{equation}

The masses of the beads in the diagonal matrix $M$ are set equal to the total mass of their corresponding cluster of atoms. The drag force at each time step is calculated using the full friction coefficient matrix obtained from the GER.

At each integration step, the laboratory-frame configuration was transformed into the corresponding Eckart body frame, where the 24 internal shape coordinates were evaluated. The generalized force was then obtained from the GMM, including the rotational-entropy correction associated with roto-vibrational coupling, in this coordinate space. These 24 generalized forces were subsequently mapped onto the
full set of Cartesian site forces by enforcing zero net intramolecular force and torque, and finally transformed back to the laboratory frame. This procedure is described in detail in \cite{rotEntropyHallGuenza}.

Importantly, no dynamical observable obtained from the CG simulations was used to fit or subsequently refine either the potential or the friction matrix. The CG trajectories therefore provide an independent validation of the dynamical model parametrized entirely from the atomistic simulations.

The coarse-grained model was used to simulate 100 replicas of duration 50ns at a timestep 1fs. The first 1ns of each trajectory was discarded as equilibration, leaving a total production simulation time of 4.9$\mu$s.

\section{Results and Discussion}

The dynamics of the atomistic and CG systems were compared through the time correlation matrix of the 24 internal shape coordinates,

\begin{equation}
C_{ij}(t) := \langle q_{i}(0)q_{j}(t)\rangle
\end{equation}

where the angular brackets denote an average over time origins and independent trajectories. The 24 internal coordinates therefore generate a complete 24 × 24 matrix containing 576 auto- and cross-correlation functions, only 300 of which are independent due to symmetry under reversal. These functions test substantially more than the relaxation of individual coordinates: the off-diagonal elements probe how fluctuations propagate and remain coupled among distinct internal motions.

\begin{figure}[t]

\begin{subfigure}{8.6cm}
\includegraphics[width = 8.6cm]{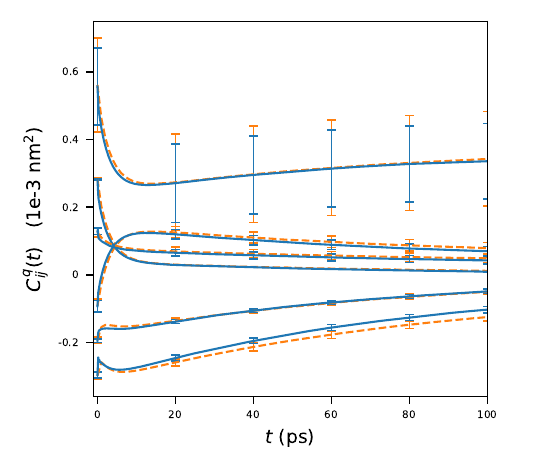} \end{subfigure}


\caption{\label{fig:mdcg_tcfs} A selection of representative comparisons between $\vec{q}$-coordinate auto- and cross-correlations in the MD (solid lines) and CG (dashed lines) models. The pairs depicted are $(i,j) \in \{(11,9), (7,16), (15,14), (0,6), (14,2), (22,1)\}$.}
\end{figure}

The correlation functions were calculated independently for each replica and subsequently averaged over replicas at each lag time. Statistical uncertainty was estimated from the replica-to-replica variation. If $\sigma_{ij} (t)$ is the standard deviation of $C_{ij} (t)$ across $N$ independent replicas, the
standard error of the mean was calculated as

$$ \sigma_{ij}^{\mathrm{mean}}(t) = \frac{\sigma_{ij}(t)}{\sqrt{N}}. $$

The error bars plotted in Fig. \ref{fig:mdcg_tcfs} represent two such standard deviations of the mean.

 Some representative comparisons between the TCFs of the MD and CG simulations are plotted in Fig. \ref{fig:mdcg_tcfs}. Plots of the full set of 300 independent conformational correlations are provided in the Supplemental Material \cite{supplementalMaterial} for further reference, and to avoid the possibility of cherry-picking. The qualitative agreement is remarkable for auto- and cross-correlations throughout the molecule, with pronounced transitions in decay rates occurring at the correct time lags and correlation amplitudes. The agreement extends to nontrivial short-time structure and to cross-correlations between coordinates with substantially different relaxation behavior.

\begin{figure}[t]
\includegraphics[width = 8.6cm]{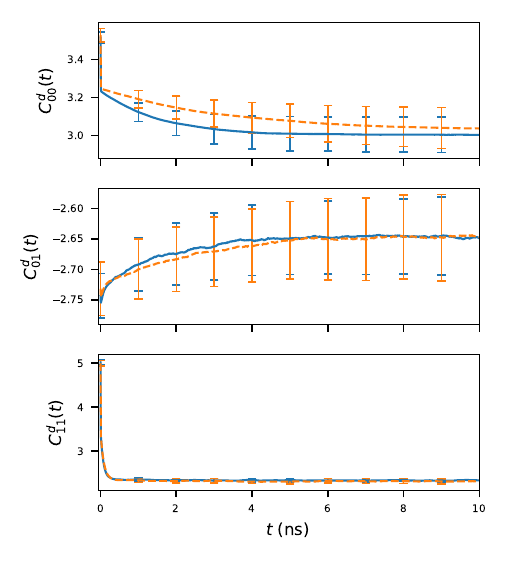}
\caption{\label{fig:dihedral_tcfs} Time-lagged correlations of the dihedral angles of alanine dipeptide, indexed as $(\phi, \psi)\rightarrow(0,1)$.}
\end{figure}

The largest systematic discrepancies occur in the longest-lived correlations, with a tendency for the most correlated motions to decay too slowly. These errors are associated with conformational transitions through regions that are poorly sampled by the unbiased atomistic trajectories. This is illustrated in Fig. \ref{fig:dihedral_tcfs} via the correlations involving the standard dihedral angles $\phi$ and $\psi$ \cite{aladi_fel_solvated, aladi_fel_solvated_2}. These angles are widely known to capture the peptide's rare transition into a metastable state, which is known to involve a barrier on the order of $~10 k_{B}T$ in solvated alanine dipeptide \cite{aladi_fel_solvated, aladi_fel_solvated_2}. The dominant basins of the GMM are well constrained, but transition rates depend sensitively on the low-probability regions between them. Consequently, relatively small errors in the multidimensional landscape near a barrier can produce appreciable errors in long-time relaxation while leaving the equilibrium marginals largely unchanged.

The present construction is deliberately based on conventional unbiased atomistic sampling. Improved sampling is therefore a natural next step, but it is distinct from the central question addressed here. The present result establishes that the equilibrium landscape and friction procedures can be combined into an independently predictive high-dimensional dynamical model. Future work can address how such models should be parameterized when rare regions cannot be adequately sampled by direct MD and, separately, how explicitly non-Markovian friction should be incorporated when the residual memory remains dynamically important.

Prior coarse-grained models of alanine dipeptide have modeled only a small number of purely internal reaction coordinates \cite{leeDarveMultidimAlanine}. Our approach is, to our knowledge, unique in its ability to accurately capture the dynamics of many internal degrees of freedom along with global translational and rotational symmetry of the molecule in a manner explicitly consistent with an underlying GLE. This work represents an advancement in the capabilities of coarse-grained models that could have applications in the study of multi-dimensional phenomena in diffusing molecules. The demonstration of significant dynamical implications of rovibrational coupling in a small peptide is also of interest, as previous studies have failed to identify rovibrational coupling in proteins using equilibrium statistics \cite{globalCouplingUbiquitin}. These tools lay groundwork for further study of such effects.

\bibliography{Bibliography}

@ARTICLE{supplementalMaterial,
   journal      = "See Supplemental Material [url]" 
}

@ARTICLE{frictionHallGuenza,
   year         = {2025},
   author       = {J. M. Hall and M. G. Guenza},
   journal      = {Phys. Rev. Lett.},
   volume       = {134},
   pages        = {147101},
}

@article{rotEntropyHallGuenza,
  title = {Coarse-grained force fields via rotational entropy corrections to free-energy landscapes of diffusing molecules},
  author = {Hall, J. M. and Guenza, M. G.},
  journal = {Phys. Rev. E},
  volume = {111},
  issue = {6},
  pages = {065304},
  numpages = {8},
  year = {2025},
  month = {Jun},
  publisher = {American Physical Society},
  doi = {10.1103/PhysRevE.111.065304},
  url = {https://link.aps.org/doi/10.1103/PhysRevE.111.065304}
}

@ARTICLE{leeDarveMultidimAlanine,
   author       = "H. S. Lee and S. Ahn and E. F. Darve",
   year         = "2019",
   journal      = "J. Chem. Phys.",
   volume       = "150",
   pages        = "174113",
}

@article{beyerleGuenza2021,
author = {Beyerle, {E.R.} and Guenza, {M.G.}},
journal = {The Journal of Chemical Physics},
volume = {154.12},
year = {2021}
}

@article{gromacs2020,
author = {Lindahl and Abraham and Hess and van der Spoel},
journal = {Zenodo},
year = {2020},
doi = {https://doi.org/10.5281/zenodo.4054996}
}

@article{amberff,
author = {Ponder, J.W. and Case, D.A.},
journal = {Adv. Prot. Chem.},
year = {2003},
volume = {66},
pages = {27-85}
}

@article{tip3p,
author = {Mark, Pekka and Nilsson, Lennart},
journal = {The Journal of Physical Chemistry A},
year = {2001},
volume = {105},
issue = {43},
pages = {9954-9960}
}

@article{aladi_fel_solvated,
author = {Smith, Paul E.},
journal = {J. Chem. Phys.},
year = {1999},
volume = {111},
pages = {5568–5579}
}

@article{aladi_fel_solvated_2,
author = {Apostolakis, Joannis and Ferrara, Philippe and Caflisch, Amadeo},
journal = {J. Chem. Phys.},
year = {1999},
volume = {110},
pages = {2099–2108}
}

@article{globalCouplingUbiquitin,
author = {Hamaneh, Mehdi Bagheri and Zhang, Liqun and Buck, Matthias},
journal = {Biophysical journal},
year = {2011},
volume = {101},
issue = {no. 1},
pages = {196-204}
}

\end{document}


\pagestyle{empty}


\title{Supplemental Information}





\author{J. M. Hall}
\email{jhall4@uoregon.edu, jmonroe.hall4@gmail.com}

 \affiliation{%
Department of Physics, University of Oregon, Eugene, Oregon 97403, USA}%

\author{M. G. Guenza}%
 \email{mguenza@uoregon.edu}

\affiliation{%
 Department of Chemistry and Biochemistry, University of Oregon, Eugene, Oregon 97403, USA
}%

\date{\today}


\maketitle

\section{Correlation Functions}

The coarse-grained (CG) model constructed here is remarkable in its ability to simultaneously model cross-correlations among many degrees of freedom with high fidelity over timescales from picoseconds to nanoseconds. For a set of 24 internal coordinates, there are 300 unique auto- and cross-correlations, given the symmetry of index reversal. Comparison plots of these 300 correlations for the MD (blue, solid) and CG (orange, dashed) simulations out to lag times of $1ns$ are included in the following pages for reference.


\clearpage


\clearpage

\newcount\pdfpage
\pdfpage=1

\loop
    \includegraphics[
        page=\the\pdfpage,
        width=\textwidth,
        height=\paperheight,
        keepaspectratio
    ]{Figures/aladi_lotsa_tcfs.pdf}
    \clearpage
    \advance\pdfpage by 1
\ifnum\pdfpage<26
\repeat